\documentclass[graphics,floatfix,amsmath,amssymb,twocolumn,aps,prl,preprintnumbers,superscriptaddress]{revtex4-2}
\usepackage{graphicx}
\usepackage{dcolumn}
\usepackage{bm}
\usepackage[utf8]{inputenc}
\usepackage[T1]{fontenc}
\usepackage{hyperref}
\usepackage{mathptmx}
\usepackage{etoolbox}
\usepackage{color}
\usepackage{natbib}
\usepackage{xcolor}
\usepackage{tikz}
\usepackage{amsmath}
\usepackage{amssymb}
\usepackage{latexsym}
\usepackage{multirow}
\usepackage{hyperref}
\usepackage[capitalise]{cleveref}
\usepackage{textcomp}
\begin{document}
\title{Enhanced dispersion of active microswimmers in confined flows}
\author{Marc Lagoin}
\thanks{marc.lagoin@u-bordeaux.fr}
\affiliation{Univ. Bordeaux, CNRS, LOMA, UMR 5798, 33405 Talence, France.}
\author{Juliette Lacherez}\thanks{These authors contributed equally to this work.}
\affiliation{Univ. Bordeaux, CNRS, LOMA, UMR 5798, 33405 Talence, France.}
\author{Guirec de Tournemire}\thanks{These authors contributed equally to this work.}
\affiliation{Univ. Bordeaux, CNRS, LOMA, UMR 5798, 33405 Talence, France.}
\author{Ahmad Badr}\thanks{These authors contributed equally to this work.}
\affiliation{Univ. Bordeaux, CNRS, LOMA, UMR 5798, 33405 Talence, France.}
\author{Yacine Amarouchene}\email{yacine.amarouchene@u-bordeaux.fr}
\affiliation{Univ. Bordeaux, CNRS, LOMA, UMR 5798, 33405 Talence, France.}
\author{Antoine Allard}\email{antoine.allard@u-bordeaux.fr}
\affiliation{Univ. Bordeaux, CNRS, LOMA, UMR 5798, 33405 Talence, France.}
\author{Thomas Salez}\email{thomas.salez@cnrs.fr}
\affiliation{Univ. Bordeaux, CNRS, LOMA, UMR 5798, 33405 Talence, France.}
\begin{abstract}
In the presence of a laminar shear flow, the diffusion of passive colloidal particles is enhanced in the direction parallel to the 
flow. This classical phenomenon is known as Taylor-Aris dispersion. Besides, microorganisms, such as active microswimmers, exhibit an effective diffusive  behavior at long times. Combining the two ingredients above, a natural question then emerges on how the effective diffusion of active microswimmers is altered in shear flows -- a widespread situation in natural environments with practical implications, \textit{e.g.} regarding biofilm formation. In this Letter, we investigate the motility and dispersion of \textit{Chlamydomonas reinhardtii} microalgae, within a rectangular microfluidic channel subjected to a sinusoidal Poiseuille flow. Using high-resolution optical microscopy and a particle-tracking algorithm, we reconstruct individual trajectories in various flow conditions and statistically analyze them through moment theory and sliding windowed demodulation. We find that the velocity fluctuations and the dispersion coefficient increase as the flow amplitude is increased, with only weak dependencies on the flow periodicity. Importantly, our results demonstrate that the generalization of Taylor-Aris law to active particles is valid. 
\end{abstract}
\maketitle

\textbf{\underline{Significance statement}. This study introduces a unified approach combining microfluidics, numerical Langevin simulations, and statistical analysis to examine how Chlamydomonas reinhardtii, an active swimming microalgae, disperses in confined shear flows. The findings extend classical Taylor-Aris dispersion theory-originally for passive Brownian particles-to active swimmers, revealing a transition from ballistic to diffusive behaviors influenced by the flow strength. Shear flows are shown to impact the swimmers' dynamics nonlinearly, enhancing spreading in complex ways. These results provide critical insight into microswimmers' behavior in fluid environments, relevant to both artificial and natural settings. The work has broader implications for microbial ecology, suggesting new ways to understand processes like biofilm formation, nutrient transport, and pathogen dispersal in ecosystems where fluid dynamics shape microorganism movement and interactions.}
\newline

Microscopic organisms frequently exceed the length scale at which thermal Brownian motion alone can mediate efficient environmental exploration. To circumvent this constraint, they employ diverse active motility strategies in viscous media~\cite{purcell2014life}, including crawling~\cite{stossel1993crawling}, gliding~\cite{mcbride2001bacterial,till2022motility}, and swimming~\cite{drescher2010direct,lisevich2025physics,hernandez2009dynamics}. Among microswimmers, propulsion mechanisms exhibit notable variations: for instance, \textit{Escherichia coli (E. coli)}  functions as a pusher, using rotary motor-driven flagella for propulsion~\cite{chattopadhyay2006swimming}; whereas the biflagellate alga \textit{Chlamydomonas reinhardtii}, from the \textit{volvox} family~\cite{hohn2015dynamics}, acts as a puller relying on synchronous beating of its flagella~\cite{drescher2010direct}. Diverse external stimuli, such as light for photosynthesis~\cite{dervaux2017light,kreis2018adhesion,de2020motility,laroussi2024short}, or chemical gradients~\cite{aubret2018targeted}, are essential in determining the dynamics of microswimmers. Similarly, hydrodynamic interactions with the environment are fundamental. As a result of these, swimming microorganisms manifest a rich repertoire of behaviors, including effective viscosity~\cite{rafai2010effective}, accumulation near boundaries~\cite{li2011accumulation,ostapenko2018curvature,thery2021rebound,souzy2022microbial}, entrainment~\cite{jeanneret2016entrainment}, upstream swimming~\cite{Brady,dey2022oscillatory}, Bretherton-Jeffery orbits~\cite{junot2019swimming}, resonant transverse swimming~\cite{hope2016resonant,guzman2012stochastic}, and emergent interactions~\cite{samatas2023hydrodynamic,grober2023unconventional}. Despite the persistent self-propelled motion of microswimmers, their long-time dynamics  often resembles the one of passive Brownian particles, which is characterized through an effective diffusion process~\cite{romanczuk2012active,solon2015active,zottl2023modeling,lowen2020inertial,saintillan2018rheology}.

In another context, the seminal studies of Taylor~\cite{taylor1953dispersion,taylor1954dispersion,taylor1954conditions} and Aris~\cite{aris1956dispersion} established that shear flows can significantly enhance the dispersion of passive microscopic solutes -- a now-classical effect known as Taylor-Aris dispersion. Further works have extended this framework to incorporate the effects of channel geometry~\cite{ajdari2006hydrodynamic}, porosity~\cite{salles1993taylor,dorfman2002generalized,brenner1980dispersion}, reactivity~\cite{shapiro1986taylor}, flow complexity~\cite{Bruus,aris1960dispersion,lee2014taylor,bearon2012biased}, short-time distribution~\cite{vilquin2021time}, and open or charged boundaries~\cite{vilquin2023nanoparticle}. The underlying coupling between advection and diffusion has remained a vibrant topic of theoretical and numerical inquiry~\cite{sadriaj2022taylor,moser2021taylor}. The Taylor-Aris mechanism has also found impactful applications, \textit{e.g.} in particle-size characterization~\cite{cottet2007taylor,cottet2010determination,cipelletti2014polydispersity,le2008size,wuelfing1999taylor}.

Given that active microswimmers effectively diffuse at long times, and often navigate within external flows in natural environnements, a compelling question arises: does an analogue of the Taylor-Aris law govern their transport? While theoretical and computational models have proliferated on active-particle dispersion in external flows~\cite{Brady,das2024taylor,kumar2021taylor,bees2010dispersion,bearon2012biased,croze2017gyrotactic}, and on the influence of oscillatory flows on active suspensions~\citep{caldag2025fine}, experimental confirmation remains scarce, primarily due to challenges in recording the dynamics over extended time scales. Nevertheless, recent and preliminary experimental studies addressed \textit{E. coli} in flows~\cite{junot_these}, bacterial scattering in microfluidic lattices~\cite{dehkharghani2019bacterial}, as well as dispersion of \textit{Chlamidomonas} as Gyrotactic swimmers~\cite{croze2017gyrotactic,caldag2025fine}. In the present Letter, we quantitatively investigate the Taylor-Aris dispersion of model active particles, using a combination of experiments, numerical simulations and theory.

We employ the green alga \textit{Chlamydomonas reinhardtii} CC409 as a model active swimmer, owing to its well-characterized motility and widespread use in microswimmer research~\cite{proschold2005portrait}. Details on cell culture and experimental preparation are provided in the Supplementary Materials (SM). Figure~\ref{figure_1} shows the experimental setup together with representative trajectories of individual cells under flow.
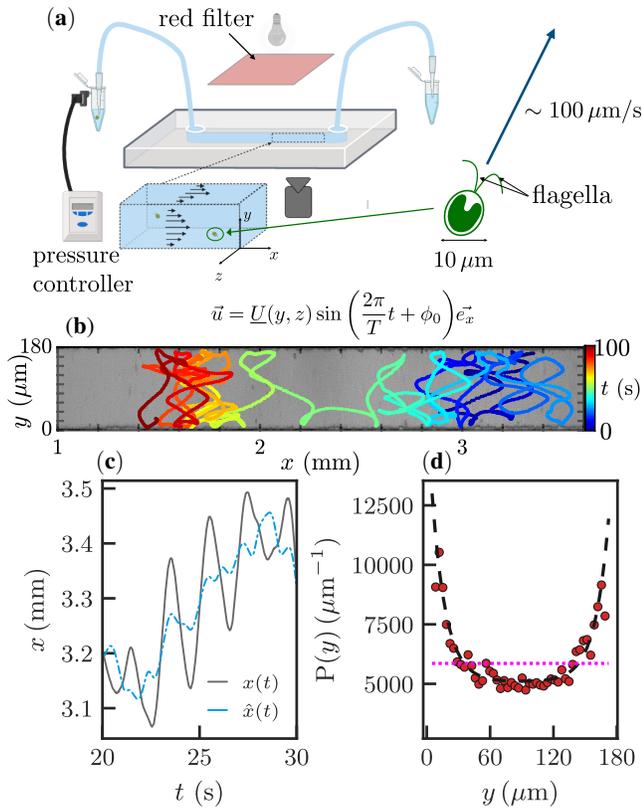
\begin{figure}[t!]
\begin{tikzpicture} 
\node(A0) at (0,0) {\includegraphics[width = 0.47\textwidth]{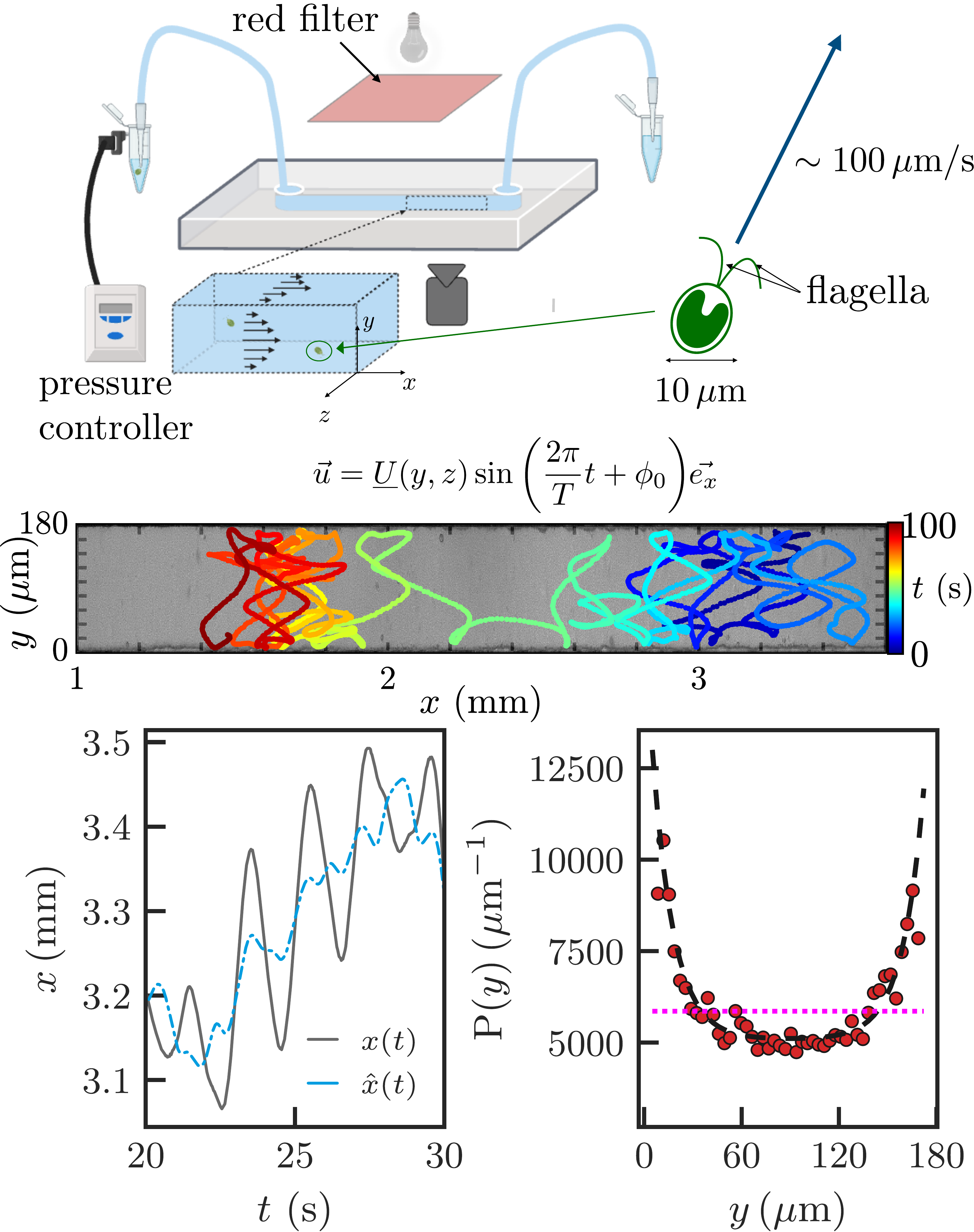}};
\node(A1) at (-3.5,5.15) {$\mathbf{(a)}$};
\node(A1) at (-3.25,1.05) {($\mathbf{b}$)};
\node(A1) at (-2.85,-0.75) {$\mathbf{(c)}$};
\node(A1) at (1.5,-0.75) {$\mathbf{(d)}$};
\end{tikzpicture} 
\caption{\textbf{Dynamics of microalgae in an oscillatory microfluidic flow.}
$\mathbf{(a)}$ Schematic of the experimental setup. A suspension of microalgae is confined in a microfluidic channel with rectangular cross-section. A pressure controller at one end generates a sinusoidal Poiseuille flow, while a constant pressure is applied at the other end via a fluid reservoir. The algae are illuminated with red-filtered light and imaged using a high-speed camera (equipped with a \(\times 5\) objective).
$\mathbf{(b)}$ Representative top-view image of the channel (\(\times 5\) magnification; frame rate: \(40\,\text{Hz}\); exposure time: \(1\,\text{ms}\)) with a typical trajectory superimposed. The colormap encodes time along the trajectory. 
$\mathbf{(c)}$ Slice of a time series of the longitudinal position \(x(t)\). The demodulated signal \(\hat{x}(t)\) is overlaid.
$\mathbf{(d)}$ Probability density function $\text{P}(y)$ of the transverse position $y$, as estimated from 27 trajectories exceeding \(20\,\text{s}\) in duration. The magenta dotted line indicates a uniform distribution. The black dashed line shows the best fit to Eq.~(\ref{Brady_Py}). In panels ($\mathbf{(b)}$,$\mathbf{(c)}$,$\mathbf{(d)}$), the experimental parameters are: P\'eclet number \(\text{Pe} = 9.15 \) (\textit{i.e.} $\underline{P}=1\,\text{mbar}$), and flow-oscillation period \(T = 2\,\text{s}\).
}\label{figure_1}
\end{figure}
This strain swims at a characteristic speed of approximately \( V_0 \approx 100\,\mu\mathrm{m}/\mathrm{s} \)~\cite{vladimirov2004measurement}, with a decorrelation time of around \(\sim1\,\mathrm{s}\), beyond which it exhibits an effective diffusion with a diffusion coefficient on the order of \( D_0 \approx 10^{-9}\,\mathrm{m}^2/\mathrm{s} \), as obtained from 1296 single-cell trajectories in the bulk and in the absence of external flow (see SM). A microfluidic channel is fabricated via soft lithography in a polydimethylsiloxane (PDMS) chip. The channel has a rectangular cross section with vertical height \( H = 100\,\mu\mathrm{m} \) along $z$, and width \( W = 180\,\mu\mathrm{m} \) along $y$, and the total length is \( L = 88\,\mathrm{mm} \) along $x$. A \(0.5\%\) w/v solution of Pluronic F-127 is used to prevent cell adhesion, following established protocols~\cite{souzy2022microbial}. The chip is filled with a dilute suspension of algae at a concentration between 2 and \( 5 \times 10^6 \) cells per millilitre, ensuring negligible cell-cell interactions. Concentrations are measured using a hemocytometer, as described in~\cite{Catalan} (see SM for further details).
To probe the long-time dynamics of active dispersion, we impose a time-periodic sinusoidal flow that maintains the algae within the field of view over a reasonable duration. Our experimental approach builds on theoretical frameworks describing active-particle dispersion in external flows~\cite{Brady} and the influence of oscillatory flows on passive thermal dispersion~\cite{Bruus}.

A pressure controller (Fluigent LU-FEZ-0345) is used to apply a target pressure \( P \) at one end of the channel. It can be modulated between \(0\) and \(345\,\mathrm{mbar}\). The controller supports sinusoidal pressure modulations with accurate waveform for periods \( T \gtrsim 0.5\,\mathrm{s} \), and guarantees precision within \(0.1\%\) of the set point for pressures above \(35\,\mathrm{mbar}\). To maintain this condition throughout the modulation cycle, the mean pressure is set to \( P_{\textrm{H}} \approx 60\,\mathrm{mbar} \), which is imposed by a hydrostatic reservoir at the opposite end of the channel. This second reservoir consists of a water column with a large free surface, and acts as a passive constant-pressure sink. Hence, the pressure is sinusoidal with mean value \( P_{\textrm{H}} \), while the oscillation amplitude \( \underline{P} \) and period \( T \) serve as independent experimental parameters. In all experiments, the pressure amplitude is set such that \( \underline{P} < 10\,\mathrm{mbar} \). The Reynolds number then satisfies $
\text{Re} = \frac{\underline{P}}{R_{\textrm{h}}} \cdot \frac{W}{\nu} \leq 0.45$,
where \( \nu \) is the kinematic viscosity of water, \( W \) is the channel width, and \( R_{\textrm{h}} \approx (3.1\pm0.2) \times 10^{5}\,\mathrm{Pa\,s/m} \) is the reduced hydraulic resistance, \textit{i.e.} the hydraulic resistance multiplied by the cross-section area $HW$, measured before any experiment. The fluid velocity field $\vec{u}(y,z,t)$ along the flow axis $x$ (with unit vector $\vec{e}_x$) can thus be accurately described as a laminar, pressure-driven Poiseuille flow, expressed as:
\begin{equation}
\vec{u}(y,z,t) = \frac{\underline{P}}{R_{\textrm{h}}}\,f(y,z) \, \sin\left( \frac{2\pi}{T} t + \phi_0 \right) \vec{e}_x\ ,
\end{equation}
where \( f(y,z) \) is the dimensionless cross-sectional profile and $\phi_0$ is an arbitrary phase shift. For instance, in a two-dimensional case, one would have \( f(y) = 4(y/W)(1 - y/W) \), while for a three-dimensional parallelepipedic channel we refer to~\cite{jonsson2009mechanical}. 
The central dimensionless control parameter in our problem is the P\'eclet number $\text{Pe}$,
that compares flow-induced advection to effective diffusion. It is defined as:
\begin{equation}
\text{Pe}\equiv \frac{\underline{P}}{\sqrt{2} R_{\textrm{h}}} \frac{W}{D_0}\ ,
\end{equation}
where \( \underline{P}/(\sqrt{2}\,R_{\textrm{h}}) \) corresponds to the root-mean-square (RMS) velocity of the oscillatory flow. Experimentally, it is not possible to exactly compensate the hydrostatic pressure, thus a small steady flow persists. The associated flow velocity is estimated from first-moment analysis on tracer trajectories to be about $10\,\mu\text{m}/\text{s}$, leading to an absolute error on the P\'{e}clet number of about $2$, which is much smaller than the range of this study.

The algae are first introduced into the microfluidic channel using a slow steady flow, before the oscillating flow is turned on. The channel is mounted on an inverted microscope equipped with a \( \times 5 \) magnification objective (Nikon CFI TU Plan Fluor BD).  \textit{Chlamydomonas} exhibit phototactic responses~\cite{Catalan,de2020motility} when illuminated with visible light, particularly in the blue-green wavelength range. To prevent this effect, they are illuminated with red-filtered light (spectrum provided in SM). Observations are made along the vertical direction $z$, averaging all motion along this direction. In practice, we track horizontal motion within a \( 4\,\mathrm{mm} \times 180\,\mu\mathrm{m} \) surface using the trackpy algorithm.
Cell trajectories are recorded for 200\,s at 40 frames per second using an Andor Neo 5.5 sCMOS camera, storing data in image stack format. Only trajectories that remain within the field of view during a reasonable time (~20\,s) are included in the analysis. For each experiment, we fix the flow period and systematically vary the pressure amplitude, and thus the P\'eclet number. A no-flow reference is recorded both before and after each experimental series. In all cases, we verified that repeated measurements do not significantly affect the motility of the algae (see SM). To minimize adaptation or memory effects between experiments, the algae are kept in the dark at zero flow for approximately 30 minutes before each new series. 
In addition, we assume that the position of a single cell can be expressed as:
\begin{equation}
    x(t) = \hat{x}(t) + A(y(t), t) \cos\left( \frac{2\pi}{T} t \right),
    \label{eq_demod}
\end{equation}
where \( A(y(t), t) \) is an unknown, potentially random function, and \( \hat{x}(t) \) represents the cell's position in the co-moving frame associated with the flow.
For a non-swimming cell, the velocity is expected to follow a sinusoidal pattern, mirroring the flow profile. If the particle is placed at a fixed transverse position \( y \), the amplitude of the oscillation is determined by the local flow velocity at that position. However, in the case of an active particle that randomly explores both \( x \) and \( y \), the amplitude of the position oscillation inherited from the flow becomes modulated by the particle's stochastic exploration along the \( y \)-axis. 

From Eq.~\ref{eq_demod}, we use sliding windowed demodulation to experimentally extract the process $\hat x(t)$. Examples of $x(t)$ and $\hat x(t)$ are shown in Fig.~\ref{figure_1}$\mathbf{(c)}$.
In Fig.~\ref{figure_1}$\mathbf{(b)}$, we observe that an individual cell explores the full channel width along $y$ multiple times within a single trajectory. This is a condition required to apply Taylor-Aris results. Furthermore, the probability density function \(\text{P}(y)\) of the transverse position \(y\), aggregated over all trajectories under identical conditions, exhibits a pronounced non-uniformity, as shown in Fig.~\ref{figure_1}$\mathbf{(d)}$. This form is well captured by the cycloidal distribution predicted for active particles~\cite{Brady}:
\begin{equation}
\label{Brady_Py}
\text{P}(y)=\mathcal{A}+ \frac{\mathcal{B}}{2}
\text{cosh} \left[\lambda\left( y- \frac{W}{2}\right)\right]
\text{, } \int\text{d} y\, \text{P}(y) =1\ ,
\end{equation}
where $\mathcal{A}$, $\mathcal{B}$ and $\lambda$ are three constants that depend on the geometry and the effective diffusion of the active process. Fit results yield: $\mathcal{A}\approx 5.1\,10^3\,\text{m}^{-1}$ which represents the bulk density, and $\mathcal{B}\approx 25\,\text{m}^{-1}$ which represents the strength of accumulation near the walls. In particular, $\lambda^{-1}\approx 13\,\mu\text{m}$, which is about the size of one cell ($7\%$ of the channel width) describes the typical size of the boundary layer close to the walls. We further assume that hydrodynamic interactions with the channel walls introduce only higher-order corrections, with axial transport primarily dictated by the bulk advective flow, in line with the Taylor–Aris dispersion framework. The validity of this assumption is demonstrated in the SM.

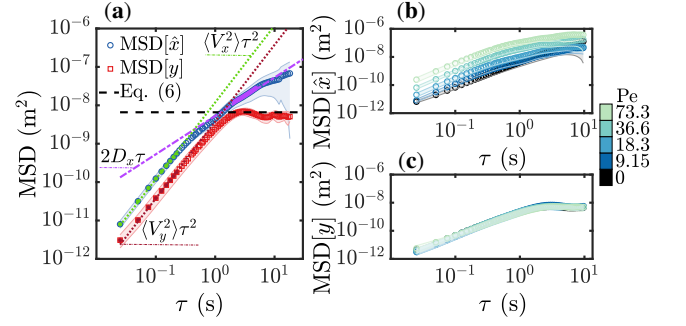
\begin{figure}[t!]
\begin{tikzpicture} 
\node(A0) at (0,0) {\includegraphics[width = 0.47\textwidth]{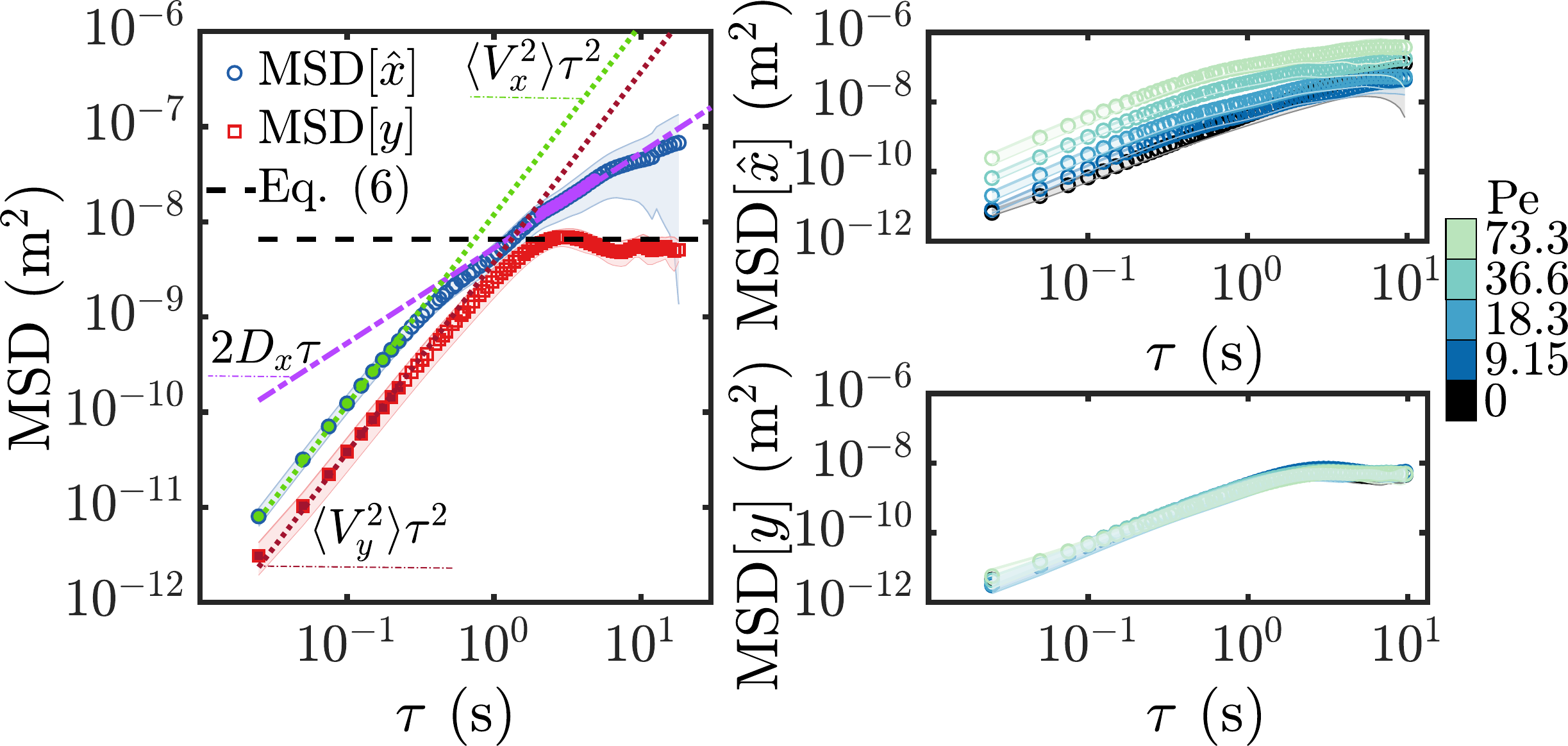}};
\node(A1) at (-2.85,2.05) {($\mathbf{a}$)};
\node(A1) at (1.05,2.05) {($\mathbf{b}$)};
\node(A1) at (1.05,0.05) {($\mathbf{c}$)};
\end{tikzpicture} 
\caption{\textbf{Mean square displacement of microalgae under oscillatory microfluidic flow.}
$\mathbf{(a)}$ Experimental ensemble-averaged mean square displacements \(\text{MSD}[\hat{x}](\tau)\) (circles) and \(\text{MSD}[y](\tau)\) (squares), respectively parallel and perpendicular to the flow, as functions of time increment $\tau$, for P\'eclet number \(\text{Pe} = 9.15\), and flow-oscillation period \(T = 2\,\text{s}\). A total of 27 trajectories, longer than 20 s each, are included in the average. Green, red, and purple symbol fillings indicate the time intervals used for estimating \(\langle V_x^2 \rangle\), \(\langle V_y^2 \rangle\), and \(D_x\), respectively, and the associated doted and dash-dotted lines indicate the ballistic and diffusive-like asymptotic behaviours, as indicated. The black dashed line corresponds to Eq.~(\ref{msdlong}), using the best-fit parameters obtained from the fit of the transverse distribution \(\text{P}(y)\) shown in Fig.~\ref{figure_1}($\mathbf{d}$). 
$\mathbf{(b)}$ Ensemble-averaged \(\text{MSD}[\hat{x}](\tau)\) for various P\'eclet numbers (see legend in next panel). 
$\mathbf{(c)}$ Ensemble-averaged \(\text{MSD}[y](\tau)\) for various P\'eclet numbers, as indicated.}
\label{figure_2}
\end{figure}

To analyse the stochastic dynamics of the cell trajectories, we consider the mean square displacements (MSDs) of the processes \(\hat{x}\) and $y$ at time scale \(\tau\). The 
MSDs are the variances of the position increments at time scale \(\tau\). For instance, for $\hat{x}$, the MSD is defined as:
\begin{equation}
\text{MSD}[\hat{x}](\tau)\equiv \overline{ \delta_\tau \hat{x}^2} - \overline{ \delta_\tau \hat{x}}^2  
 \text{, } \delta_\tau \hat{x}(t)\equiv \hat{x}(t+\tau)-\hat{x}(t)\ ,
\end{equation}
where $\overline{ \; \textcolor{white}{\cdot} \; }$ denotes the temporal average. In Fig.~\ref{figure_2}($\mathbf{a}$), we present typical experimental measurements of \(\mathrm{MSD}[\hat{x}]\) and \(\mathrm{MSD}[y]\) for fixed flow conditions. 
Asymptotically, for time scales much smaller than the decorrelation time \(\tau_x \approx 1\,\mathrm{s}\) in the $x$ direction, the parallel MSD exhibits a ballistic scaling, \textit{i.e.} $
\mathrm{MSD}[\hat{x}](\tau \ll \tau_x) \simeq \langle V_x^2 \rangle \tau^2$ ,
where \(\langle V_x^2 \rangle\) is the mean square velocity. Conversely, for time scales much larger than \(\tau_x\), the dynamics is diffusive-like, \textit{i.e.} $
\mathrm{MSD}[\hat{x}](\tau \gg \tau_x) \simeq 2 D_x \tau$, 
where \(D_x\) is the so-called dispersion coefficient. The latter coefficient encompasses in a general fashion the effects of: i) the raw effective diffusion due to activity; and ii) the possible Taylor-Aris mechanism at stake in this study. Besides, $\text{MSD}[y]$ also exhibits a ballistic behaviour at short times, meaning that the algae swims both parallel and perpendicular to the flow. However, we cannot observe a long-time diffusive-like behaviour perpendicular to the flow, as $\text{MSD}[y]$ saturates to a plateau at large times due to the presence of the walls, which impose $y\in [0,\,W]$. Indeed, we find good agreement when comparing the plateau value of $\text{MSD}[y]$ with the prediction:
\begin{equation}
\label{msdlong}
\text{MSD}[y](\tau \longrightarrow + \infty )=2 \int\text{d}y\, y^2 \text{P}(y)\ .
\end{equation} 
Moreover, in Fig.~\ref{figure_2}($\mathbf{b}$),  we observe that $\text{MSD}[\hat{x}]$ increases as $\text{Pe}$ increases, which indicates that both the variance of the speed and the dispersion coefficient grow with the P\' eclet number. In sharp contrast, this behaviour is absent for $\text{MSD}[y]$, as all the curves  of Fig.~\ref{figure_2}($\mathbf{c}$) collapse together for any value of $\text{Pe}$. We conclude from this latter observation that in our experimental range of $\text{Pe}$, the flow does not disturb significantly the algae swimming mechanism itself.

We now focus on the evolution of the speed variance (see details in SM) with the flow amplitude. 
\begin{figure}[t!]
\begin{tikzpicture} 
\node(A0) at (0,0) {\includegraphics[width = 0.47\textwidth]{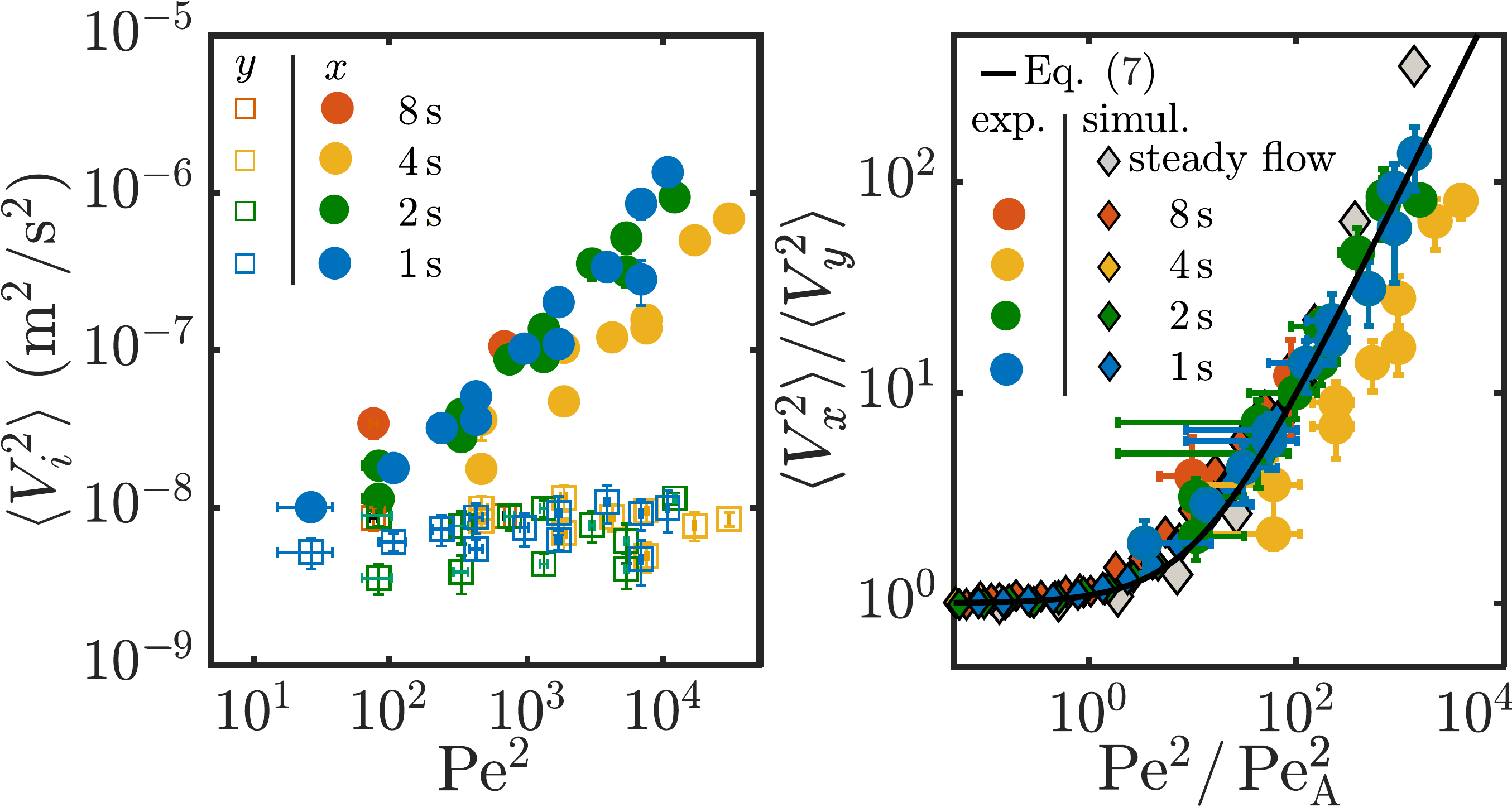}};
\node(A1) at (-2.85,2.25) {($\mathbf{a}$)};
\node(A1) at (1.35,2.25) {($\mathbf{b}$)};
\end{tikzpicture} 
\caption{
\textbf{Velocity fluctuations  of microalgae under oscillatory microfluidic flow.}
$\mathbf{(a)}$ Experimental ensemble-averaged mean square velocities \(\langle V_x^2 \rangle\) (circles) and \(\langle V_y^2 \rangle\) (squares), respectively parallel and perpendicular to the flow, as functions of the squared P\'eclet number \(\text{Pe}^2\), for various flow-oscillation periods $T$, as indicated. 
$\mathbf{(b)}$ Ratio of the mean square velocities \(\langle V_x^2 \rangle / \langle V_y^2 \rangle\) as a function of the squared ratio between P\'eclet number \(\text{Pe}\) and active P\'eclet number~\cite{romanczuk2012active,solon2015active,zottl2023modeling,lowen2020inertial,saintillan2018rheology} $\text{Pe}_{\mathrm{A}}\equiv (W/D_0)\sqrt{\langle V_y^2 \rangle}\approx 3.2$, for various flow-oscillation periods $T$, as indicated. The circles correspond to experimental data while the diamonds correspond to the results from two-dimensional Langevin-like simulations~\cite{Brady} in oscillatory flows, including one steady-flow reference case.
The solid black line corresponds to Eq.~(\ref{prediction_v}) with $\beta=8/90$ (see SM). 
Each point represents an ensemble average over trajectories longer than \(20\,\mathrm{s}\). Vertical error bars denote one standard deviation, while horizontal error bars account for uncertainties in the applied pressure (as specified by the manufacturer) and in the measurement of the baseline diffusion coefficient \(D_0\) (see SM).
}\label{figure_3}
\end{figure}
In Fig.~\ref{figure_3}$\mathbf{(a)}$, we plot the experimental \(\langle V_x^2 \rangle\) and \(\langle V_y^2 \rangle\) as functions of $\text{Pe}^2$, for different flow-oscillation periods $T$. Our results indicate that the flow-oscillation period has no significant effect within our parameter range. This is expected since, by design of our study, the period of the oscillation was chosen to remain greater than the typical decorrelation time of the particle random motion~\cite{Bruus}. 
Also, we recover that $\langle V_y^2 \rangle$ remains unchanged by the flow amplitude. Interestingly, we observe that $\langle V_x^2 \rangle$ increases linearly with $\text{Pe}^2$, which is consistent with speed composition. Indeed, for a Brownian particle in a two-dimensional Poiseuille flow with maximum speed $\underline{U}$, one can prove that $\langle V_x^2 \rangle / \langle V_y^2 \rangle = 1+ \beta  \underline{U}^2  /\langle V_y^2 \rangle $, where $\beta=8/90$ is a geometric factor (see SM). Assuming that this analysis is valid for active particles at long times, in our three-dimensional configuration, and using the RMS flow speed relevant to our oscillatory case instead of $\underline{U}$, one obtains: 
\begin{equation}
\label{prediction_v}
\frac{ \langle V_x^2 \rangle }{\langle V_y^2 \rangle } = 1 + \beta  \frac{    \text{Pe}^2   }{\text{Pe}_{\mathrm{A}}^{\,2}}\ ,
\end{equation}
where we have introduced the usual, zero-flow, active P\'eclet number~\cite{romanczuk2012active,solon2015active,zottl2023modeling,lowen2020inertial,saintillan2018rheology}, $\text{Pe}_{\mathrm{A}}\equiv (W/D_0)\sqrt{\langle V_y^2 \rangle}\approx 3.2$. In addition, we perform classical Langevin-like numerical simulations of active particles~\cite{Brady} in two dimensions, using a forward-Euler discretization scheme, but with the inclusion of an oscillatory Poiseuille flow. In Fig.~\ref{figure_3}$\mathbf{(b)}$, we confront the prediction of Eq.~(\ref{prediction_v}) to both the experimental and numerical data. We observe good agreement between theory, simulations, and experiments. 

\begin{figure}[t!]
\begin{tikzpicture} 
\node(A0) at (0,0) {\includegraphics[width = 0.47\textwidth]{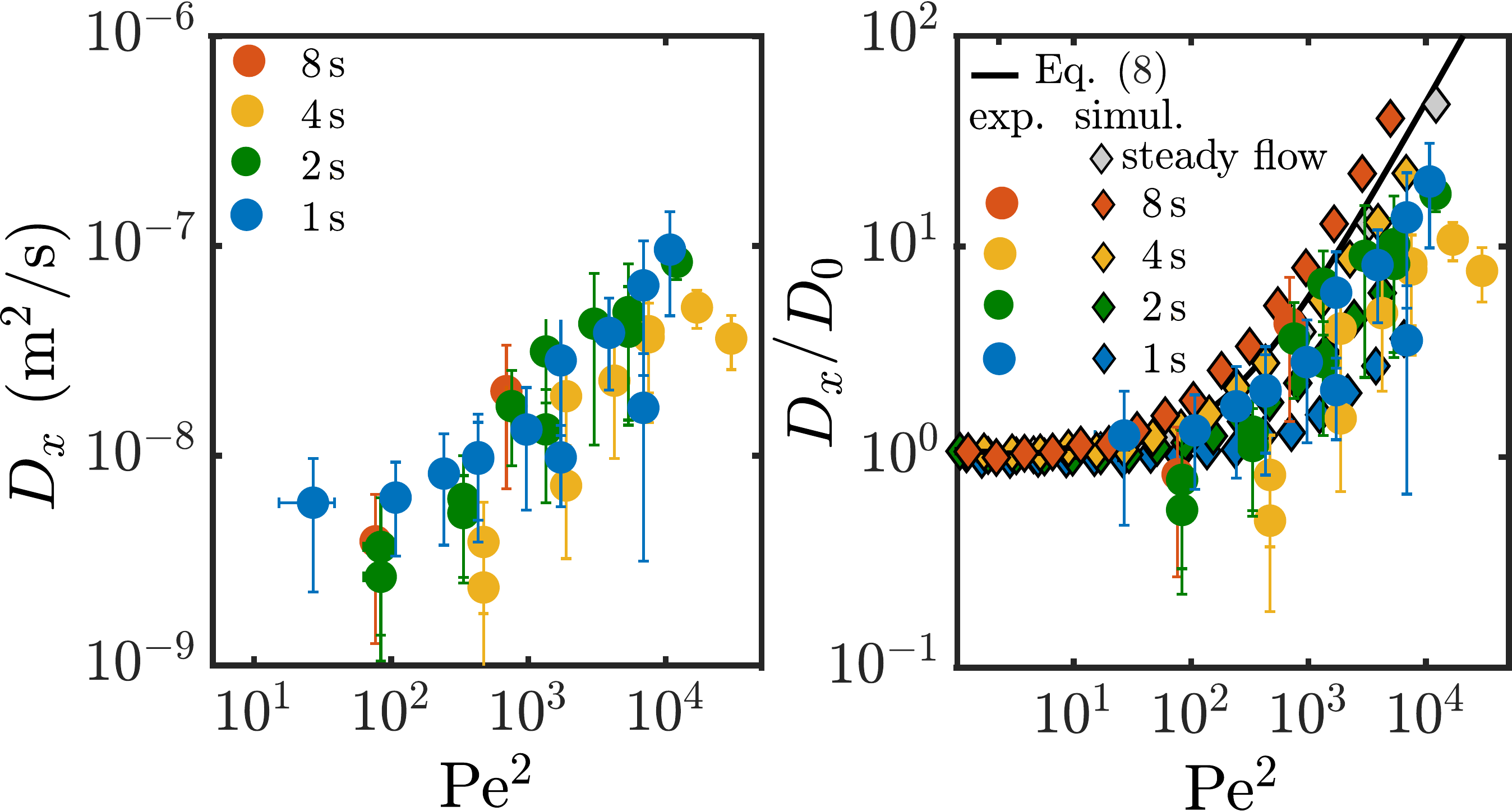}};
\node(A1) at (-2.85,2.25) {($\mathbf{a}$)};
\node(A1) at (1.35,2.25) {($\mathbf{b}$)};
\end{tikzpicture} 
\caption{\textbf{Dispersion of microalgae under oscillatory microfluidic flow.}
$\mathbf{(a)}$ Experimental ensemble-averaged dispersion coefficient $D_x$ parallel to the flow as a function of the squared P\'eclet number \(\text{Pe}^2\), for various flow-oscillation periods $T$, as indicated. 
$\mathbf{(b)}$ Ratio $D_x/D_0$ of the parallel dispersion coefficient to the raw effective diffusion coefficient  as a function of squared P\'eclet number \(\text{Pe}^2\), for various flow-oscillation periods $T$, as indicated. The circles correspond to experimental data while the diamonds correspond to the results from two-dimensional Langevin-like simulations~\cite{Brady} in oscillatory flows, including one steady-flow reference case.
The solid black line corresponds to Eq.~(\ref{TA_ideal}) with $\alpha=1/210$. 
Each point represents an ensemble average over trajectories longer than \(20\,\mathrm{s}\). Vertical error bars denote one standard deviation, while horizontal error bars account for uncertainties in the applied pressure (as specified by the manufacturer) and in the measurement of the baseline diffusion coefficient \(D_0\) (see SM).}
\label{figure_4}
\end{figure}

Let us finally investigate long-term dispersion. In Fig.~\ref{figure_4}$\mathbf{(a)}$, we show the variation of the experimental dispersion coefficient \(D_x\) parallel to the flow direction as a function of the flow amplitude.
We see no major systematic effect of the flow-oscillation period beyond error bars. Interestingly, we observe that \(D_x\) increases linearly with \(\mathrm{Pe}^2\), which is consistent with the classical Taylor-Aris law for passive Brownian particles in steady viscous shear flows~\cite{taylor1953dispersion,taylor1954dispersion,taylor1954conditions,aris1956dispersion}:
\begin{equation}
\label{TA_ideal}
D_x=D_0\left(1+\alpha \text{Pe}^2 \right)\ ,
\end{equation}
where $\alpha$ is a geometric factor, that is equal to $1/210$ in two-dimensional Poiseuille flows, and where $D_0$ denotes the bulk diffusion coefficient of the considered particles. To test this mechanism further, we plot in Fig.~\ref{figure_4}$\mathbf{(b)}$ the ratio of the parallel dispersion coefficient $D_x$ to the raw effective diffusion coefficient $D_0$ as a function of the squared P\'eclet number \(\mathrm{Pe}^2\), for all the experimental and numerical data.
The data is consistent with Eq.~(\ref{TA_ideal}), hence revealing the main finding that the classical Taylor-Aris mechanism for dispersion of Brownian particles in shear flows can be extended to the case of model active swimmers exhibiting long-term effective diffusion. Two minor remarks should however be made. As a first remark, we measure \(\alpha \equiv (D_x/D_0 - 1)/\mathrm{Pe}^2\approx1.6\pm2.3\,10^{-3}\) from the experimental data, which, while being on the same order of magnitude, is systematically below the theoretical value $\alpha=1/210\approx 4.8\,10^{-3}$. This discrepancy may have several origins, beyond the obvious dimensionality difference which provides only minor corrections. First, by construction, long-time experimental data is more prone to statistical errors. Second, the cells may adapt to the flow in hardly predictable and testable ways. Third, $D_x$ and $D_0$ are expected to conjointly vary from one living organism to the other, which implies that both should in principle be simultaneously measured for each given individual cell. Unfortunately, it was not experimentally feasible to ensure that no cell entered or left the channel during one single experiment. It was therefore not possible to obtain the $D_0$ value corresponding to each $D_x$ measurement. Instead, as already mentioned, we use the average effective diffusion coefficient measured on many algae, in the bulk, and in the absence of any flow (see SM). As a second remark, we notice that the simulation results show a dependency with the flow period, that the experiments cannot confirm due to the current statistical resolution. Reducing this uncertainty would likely require an alternative experimental approach, such as tracking individual cells over multiple flow cycles and applying appropriate normalization to each trajectory.

In this Letter, we combined microfluidic experiments on \textit{Chlamydomonas reinhardtii} microalgae, numerical Langevin simulations, and theory, in order to address the dispersion of model active microswimmers in viscous shear flows. Combining single-cell tracking and statistical analysis, we checked the expectation that algae preferentially accumulate near channel boundaries. Furthermore, in the direction parallel to the flow, we observed and quantified a ballistic-to-diffusive-like crossover, with a clear influence of the flow amplitude in both asymptotic regimes. We quantitatively rationalized all the parametric dependencies. Our central finding is that the classical Taylor-Aris law for dispersion of Brownian particles in shear flows can be extended to the case of model active microswimmers. This mechanism could have important practical implications in microorganism spreading and biofilm formation, within natural environnements where flows are omnipresent. In future, it would be interesting to test the mechanism for model pushers rather than pullers, to incorporate external stimuli such as light or chemical cues, in order to bias the active Taylor-Aris dispersion uncovered here, and to test additional lubrication effects in stronger confinement. Moreover, we here used quasi-steady oscillatory flows to ensure sufficient statistical resolution, but increasing further the oscillation frequency beyond the natural decorrelation rate~\cite{Bruus} and investigating the resulting biomechanical change in swimming would be an interesting avenue.

 \section*{Data availability}
Videos and numerical routines can be found on the online repository: \url{https://github.com/EMetBrown-Lab/Chlamy_taylor_aris/tree/main}.

 \section*{Acknowledgements}
The authors thank Françoise Argoul, Robin Dumora Beyriere, Nicolas Fares, Quentin Ferreira, Aurelia Honerkamp-Smith, Gaspard Junot, Maxime Lavaud, Marco Polin, St\'ephane G. Roux, and Alexandre Vilquin for interesting discussions, as well as Elodie Millan for help with figure design. They acknowledge financial support from the European Union through the European Research Council under EMetBrown (ERC-CoG-101039103) grant. They also acknowledge financial support from the Agence Nationale de la Recherche under EMetBrown (ANR-21-ERCC-0010-01), Softer (ANR21-CE06-0029), and Fricolas (ANR-21-CE06-0039) grants, as well as from the Interdisciplinary and Exploratory Research program under MISTIC grant at University of Bordeaux, France. Besides, they also acknowledge the support from the LIGHT Sciences and Technologies Graduate Program (PIA3 Investment for the Future Program, ANR-17EURE-0027), the Matter and Light Science Department at University of Bordeaux, the Tremplin INP CNRS program, and the Réseaux de Recherche Impulsion (RRI) “Frontiers of Life”, which received financial support from the French government in the framework of the University of Bordeaux’s France 2030 program. 
Finally, they thank the Soft Matter Collaborative Research Unit, Frontier Research Center for Advanced Material and Life Science, Faculty of Advanced Life Science at Hokkaido University, Sapporo, Japan, and the CNRS International Research Network between France and India on ``Hydrodynamics at small scales: from soft matter to bioengineering.
\bibliographystyle{ieeetr}
\bibliography{Lagoin2025}
\end{document}